\documentclass[10pt,twocolumn,letterpaper]{article}

\usepackage{iccv}
\usepackage{times}
\usepackage{epsfig}
\usepackage{graphicx}
\usepackage{amsmath}
\usepackage{amssymb}
\usepackage[dvipsnames,svgnames]{xcolor}
\usepackage[pagebackref=true,breaklinks=true,letterpaper=true,colorlinks,bookmarks=false,linkcolor=red,anchorcolor=blue,citecolor=DodgerBlue]{hyperref}
\usepackage{multirow}
\usepackage[accsupp]{axessibility}  

\iccvfinalcopy 

\def\iccvPaperID{2619} 

\ificcvfinal\fi

\begin{document}

\title{CuNeRF: Cube-Based Neural Radiance Field for Zero-Shot Medical Image Arbitrary-Scale Super Resolution}

\author{Zixuan Chen$^{1}$\qquad Lingxiao Yang$^{1,2,3}$\qquad Jian-Huang Lai$^{1,2,3}$\qquad Xiaohua Xie$^{1,2,3}$\thanks{Corresponding author.}\\
$^1$School of Computer Science and Engineering, Sun Yat-sen University, Guangzhou, China\\
$^2$Guangdong Province Key Laboratory of Information Security Technology, Guangzhou, China\\
$^3$Key Laboratory of Machine Intelligence and Advanced Computing, Ministry of Education, China\\
\small{\ttfamily{chenzx3@mail2.sysu.edu.cn, lingxiao.yang717@gmail.com, \{stsljh, xiexiaoh6\}@mail.sysu.edu.cn}}
}

\maketitle
\ificcvfinal\thispagestyle{empty}\fi

\newcommand{\bt}[1]{\textbf{#1}}
\newcommand{\ul}[1]{\underline{#1}}
\newcommand{\ck}{\checkmark}
\newcommand{\ua}{$\uparrow$}
\newcommand{\da}{$\downarrow$}
\newcommand{\mc}[1]{\multicolumn{2}{|c}{#1}}
\newcommand{\mr}[1]{\multirow{2}{*}{#1}}

\begin{abstract}
  Medical image arbitrary-scale super-resolution (MIASSR) has recently gained widespread attention, aiming to supersample medical volumes at arbitrary scales via a single model.
  However, existing MIASSR methods face two major limitations: \bt{(i)} reliance on high-resolution (HR) volumes and \bt{(ii)} limited generalization ability, which restricts their applications in various scenarios.
  To overcome these limitations, we propose Cube-based Neural Radiance Field (CuNeRF), a zero-shot MIASSR framework that is able to yield medical images at arbitrary scales and free viewpoints in a continuous domain.
  Unlike existing MISR methods that only fit the mapping between low-resolution (LR) and HR volumes, CuNeRF focuses on building a continuous volumetric representation from each LR volume without the knowledge of the corresponding HR one. 
  This is achieved by the proposed differentiable modules: cube-based sampling, isotropic volume rendering, and cube-based hierarchical rendering.
  Through extensive experiments on magnetic resource imaging (MRI) and computed tomography (CT) modalities, we demonstrate that CuNeRF can synthesize high-quality SR medical images, which outperforms state-of-the-art MISR methods, achieving better visual verisimilitude and fewer objectionable artifacts. 
  Compared to existing MISR methods, our CuNeRF is more applicable in practice.
\end{abstract}

\begin{figure*}[!t]
  \centering
  \includegraphics[width=\textwidth]{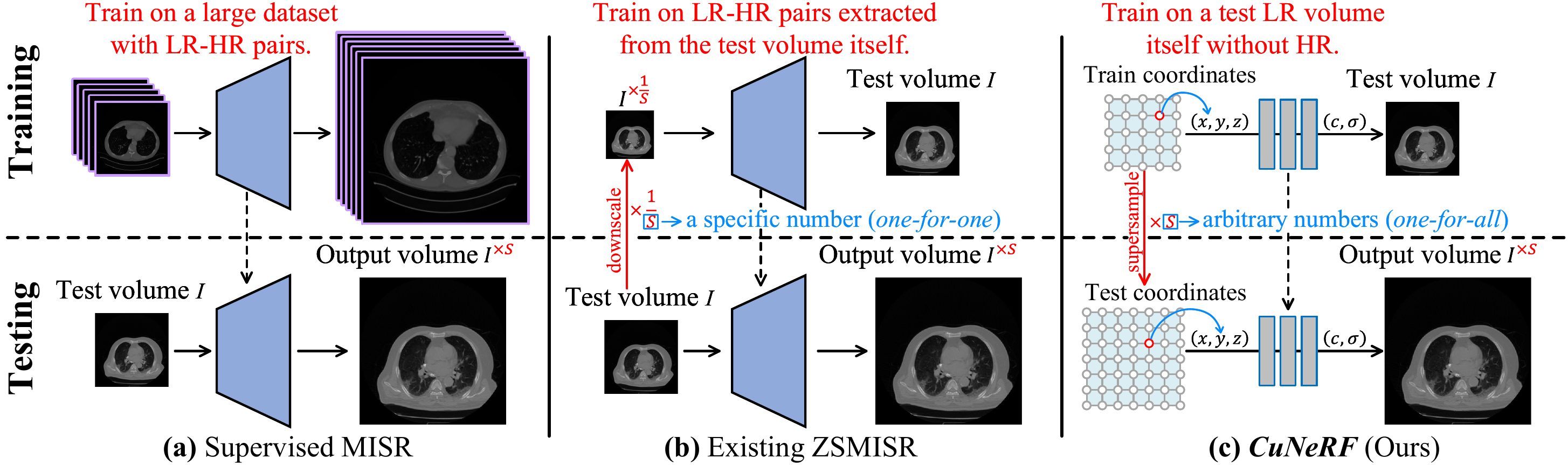}
  \caption{
    Difference between existing supervised MISR \bt{(a)}, zero-shot MISR (ZSMISR) \bt{(b)} and \textit{CuNeRF} \bt{(c)}.
    Visually, supervised MISR methods need to collect considerable LR-HR pairs for training, while ZSMISR and our \textit{CuNeRF} only train the model on each test volume itself.
    However, given a test volume, ZSMISR methods can only upsample medical images at a specific scale (one-for-one), while our \textit{CuNeRF} can handle arbitrary upsampling scales (one-for-all). 
  }
  \label{fig:zssr}
\end{figure*}

\section{Introduction}\label{sec:intro}
Medical imaging techniques such as computed tomography (CT) and magnetic resonance imaging (MRI) are critical tools in assisting clinical diagnosis.
However, the acquisition of high-quality medical slices is a resource-intensive process, which requires subjects to be exposed to considerable ionizing radiations for a long time, increasing the lifetime risk of cancer \cite{CT}.
To reduce the burden on subjects, a feasible approach is to reconstruct high-resolution (HR) medical volumes from low-resolution (LR) ones.

To tackle medical image super-resolution (MISR) challenges, early studies employed optimization methods \cite{MISR,CSLC} and interpolation methods \cite{tricubic}.
Subsequently, a series of methods \cite{ssp,ctsrgan,csn,gan3dmdcn,egan} have adopted convolutional neural networks to learn the LR-HR mappings.
Recently, medical image arbitrary-scale super-resolution (MIASSR) methods \cite{saint,ARSSR} have received widespread attention in the MISR community, aiming to employ a single model to upsample medical volumes at arbitrary scales.
Although these methods achieve acceptable HR results, they still have two major issues: \textit{(i)} Existing MIASSR methods rely on the supervision from HR volumes, yet high-quality HR volumes are not always available;
\textit{(ii)} These methods may be susceptible to the distribution gap between training and test data, producing non-existent details.
These drawbacks limit the application scenarios of existing MIASSR methods.

To address the above-mentioned limitations, we present a zero-shot MIASSR framework -- Cube-based NeRF (CuNeRF), which aims to yield arbitrary upsampling images after training on a test LR volume itself (see Figure \ref{fig:zssr}).
Specifically, we draw inspiration from the neural radiance field (NeRF) \cite{NeRF} to estimate the continuous volumetric representation from discrete samples (LR volumes) instead of fitting the mapping between LR and HR volumes.
Since directly applying NeRF on medical volumes may result in grid-like artifacts (see Figure \ref{fig:examples}, detailed explanation is provided in Section \ref{sec:holes}), our \textit{CuNeRF} tackles such aliasing issues via the proposed differentiable modules: cube-based sampling, isotropic volume rendering, and cube-based hierarchical rendering.
As shown in Figure \ref{fig:contribution}, \textit{CuNeRF} can build a continuous mapping between the coordinate and the corresponding intensity value in the training data, which is capable of generating medical slices at arbitrary scales and free viewpoints in a continuous domain.
Comprehensive experiments on the MSD Brain Tumour (MRI) \cite{MSD} and KiTS19 (CT) \cite{kits19} datasets show that \textit{CuNeRF} yields impressive performance in 3D and volumetric MISR at various upsampling scales, outperforming state-of-the-art methods.

The main contributions are summarized:

\begin{itemize}
  \item To the best of our knowledge, \textit{CuNeRF} is the first zero-shot MIASSR framework that can continuously upsample medical volumes at arbitrary scales.
  
  \item We address the hole-forming issues via the proposed techniques: cube-based sampling, isotropic volume rendering, and cube-based hierarchical rendering.

  \item Extensive experiments on CT and MRI modalities for 3D MISR and volumetric MISR show \textit{CuNeRF} favorably surpasses state-of-the-art MIASSR methods. 
  
\end{itemize}   
   
\section{Related Works}
In this section, we first review implicit neural representation and then introduce some impressive progress in medical image super-resolution.
Recent surveys \cite{survey,msurvey} provide a comprehensive review of super-resolution methods.

\subsection{Implicit Neural Representation}
Learning implicit neural representations (INRs) from discrete samples to form a continuous function has been a long-standing research problem in computer vision for numerous tasks.
A recent trend in this field is to map discrete representations to coordinate-based continuous neural representations through implicit functions formed by neural networks, such as multi-layer perceptron (MLP).
Chen \emph{et al.} \cite{LIIF} proposed a method to learn the INR of 2D images using the local implicit image function. Subsequent work \cite{videoinr} extended this \cite{LIIF} to apply in the video domain.
Currently, most 3D view-synthesis methods are based on the neural radiance fields (NeRF) \cite{NeRF} framework.
NeRF can model a volumetric radiance field to render novel views with impressive visual quality using standard volumetric rendering \cite{rendering} and alpha compositing techniques \cite{alpha}.
However, NeRF has the drawback of requiring massive training views and lengthy optimization iterations to learn the correct 3D geometry.
Several follow-up works have attempted to optimize NeRF's training procedures, such as reducing the required training views \cite{pixnerf,dnerf,Sparsenerf}, accelerating convergence and rendering speed \cite{instngp}.
Other works aim to adapt NeRF to various domains, such as generative modeling \cite{GRAF,Giraffe}, anti-aliasing \cite{mip-nerf}, unbounded representation \cite{mip-nerf-360}, and RGB-D scene synthesis \cite{rgbd-nerf}.
Recently, some researchers employ INR-based methods to reconstruct medical images from discrete-sampled data \cite{GRFF,IntraTomo,APRF,mednerf}, more details can be seen in the recent survey \cite{INR4MedicalSurvey}.

\subsection{Medical Image Super Resolution}\label{sec:related}
Medical image super-resolution (MISR) is an important task in medical image processing, which aims to reconstruct high-resolution (HR) medical slices from corresponding low-resolution (LR) ones.
Initially, some conventional methods like \cite{MISR,CSLC} and widely-used interpolation methods like bicubic and tricubic interpolations \cite{tricubic} were employed in the early research.
Inspired by \cite{dlsr}, recent studies have shifted their focus towards using deep learning-based super-resolution networks in the medical domain. 
Lim \emph{et al.} \cite{edsr} employ deep learning-based super-resolution networks to upsample medical images.
Some studies upsample each 2D LR medical slice to acquire the corresponding HR one, such as \cite{ssp,ctsrgan,csn}. On the other hand, Chen \emph{et al.} \cite{gan3dmdcn} and Wang \emph{et al.} \cite{egan} use 3D DenseNet-based networks to generate HR volumetric patches from LR ones.
Yu \emph{et al.} \cite{tvsrn} build a transformer-based MISR network to address volumetric MISR challenges.
Recent studies have been focusing on medical image arbitrary-scale super-resolution (MIASSR) \cite{ARSSR,saint}, which aims to upsample medical slices at arbitrary scales by a single model.
Inspired by Meta-SR \cite{meta-sr}, Peng \emph{et al.} \cite{saint} deals with volumetric MISR on the $z$-axis at integer scales.
Wu \emph{et al.} \cite{ARSSR} propose ArSSR, an INR-based method that can upsample MRI volumes at arbitrary scales in a continuous domain.
Wang \emph{et al.} \cite{uassr} propose a weakly-supervised framework that uses unpaired LR-HR medical volumes for optimization.
However, these methods deeply rely on the HR medical volumes, which limits the application scenarios.
   
\section{Preliminary: NeRF}\label{sec:NeRF}
Neural radiance field (NeRF) \cite{NeRF} aims to build the continuous mapping from $(\mathbf{x}, \mathbf{d})$ to $(\mathbf{c}, \sigma)$, where $\mathbf{x}=(x,y,z)$ and $\mathbf{d}=(\theta, \phi)$ denote spatial location and viewing direction, while $\mathbf{c}$ and $\sigma$ represent the content color and volume density, respectively.
NeRF's techniques can be summarized as follow:

\noindent\bt{Ray sampling.} NeRF first constructs the ray $\mathbf{r}(t)=\mathbf{o}+t\mathbf{d}$ that emits from the center of projection $\mathbf{o}$ and passes through the materials along the viewing direction $\mathbf{d}$.
Subsequently, NeRF samples $N$ points along the ray from near plane $t_{n}$ to far plane $t_{f}$ predefined.
For each sampling point $\mathbf{r}(t_{k})$, NeRF employs a positional encoding function $\gamma(\cdot)$ to map the location $\mathbf{x}_{k}$ and view direction $\mathbf{d}$ into higher dimensional space as:
\begin{equation}
  \gamma(\rho)=\rho\bigcup^{L-1}_{i=0}(\sin(2^{i}\rho), \cos(2^{i}\rho)),\ where\ L\in\mathbb{N}.
  \label{eq:PE}
\end{equation}
where $\rho$ denotes an arbitrary vector and $L$ is a hyperparameter set to $10$ as default.

\noindent\bt{Volume rendering.} The pixel color $\mathbf{C}(\mathbf{r})$ can be modeled as the integral of the corresponding ray $\mathbf{r}$ based on Beer-Lambert Laws as:
\begin{equation}
  \mathbf{C}(\mathbf{r}) = \int_{t_{n}}^{t_{f}}\frac{\sigma(\mathbf{r}(t))\mathbf{c}(\mathbf{r}(t),\mathbf{d})dt}{\exp(\int_{t_{n}}^{t}\sigma(\mathbf{r}(s))ds)},
  \label{eq:integral}
\end{equation}
where $\mathbf{c}(\cdot)$ and $\sigma(\cdot)$ denote the color and volume density functions.
In practice, NeRF employs a multi-layer perceptron (MLP) $F_{\Theta}$ to estimate these two functions.
For each sampling point $\mathbf{r}(t_{k})$, MLP $F_{\Theta}$ predicts the corresponding color $\mathbf{c}_{k}$ and volume density $\sigma_{k}$ by:
\begin{equation}
(\mathbf{c}_{k}, \sigma_{k}) = F_{\Theta}(\gamma(\mathbf{x}_{k}), \gamma(\mathbf{d})).
\label{eq:MLP}
\end{equation}
Given the estimated results of the $N$ sampling points from $t_{n}$ to $t_{f}$, we can approximate the volume rendering integral using numerical quadrature as introduced by \cite{Max}:
\begin{equation}
  \hat{\mathbf{C}}(\mathbf{r})=\sum^{N}_{i=1}\frac{1-\exp(-\sigma_{i}(t_{i+1}-t_{i}))}{\exp(\sum_{j=1}^{i}\sigma_{j}(t_{j+1}-t_{j}))}\mathbf{c}_{i},
  \label{eq:volume}
\end{equation}
where $\hat{\mathbf{C}}(\mathbf{r})$ is the predicted color of the pixel.

\noindent\bt{Hierarchical volume rendering.} NeRF also refine the result by allocating samples proportionally to their expected volume distribution based on the coarse estimations.
NeRF simultaneously optimizes two MLPs, \emph{i.e.}, the coarse one $F^{c}_{\Theta}$ and the fine one $F^{f}_{\Theta}$.
Specifically, NeRF first samples $N_{c}$ points and obtain the coarse output $\hat{\mathbf{C}}_{c}(\mathbf{r})$ by Eq \ref{eq:volume}, which can be rewrited as $\hat{\mathbf{C}}_{c}(\mathbf{r})=\sum^{N_{c}}_{i=1}w_{i}\mathbf{c}_{i}$.
A piecewise-constant PDF related to the sampling points along the ray can be produced by $\hat{w} = w_{i}/\sum^{N_{c}}_{j=1}w_{j}$.
NeRF then samples $N_{f}$ points from this distribution by inverse transform sampling (ITS) and computes the fine outputs $\hat{\mathbf{C}}_{f}(r)$ using all $N_{c}+N_{f}$ sorted sampling points.
Let $\mathcal{R}$ represent the batch, and these two MLPs can be optimized by the following rendering loss:
\begin{equation}
\mathcal{L}=\sum_{\mathbf{r}\in \mathcal{R}}\left[\|g.t. - \hat{\mathbf{C}}_{c}(\mathbf{r})\|^{2}_{2}+\|g.t. - \hat{\mathbf{C}}_{f}(\mathbf{r})\|^{2}_{2}\right],
\end{equation}
where $g.t.$ denotes the ground truth of the rendering pixels.

\begin{figure}[!t]
  \centering
  \includegraphics[width=3.2in]{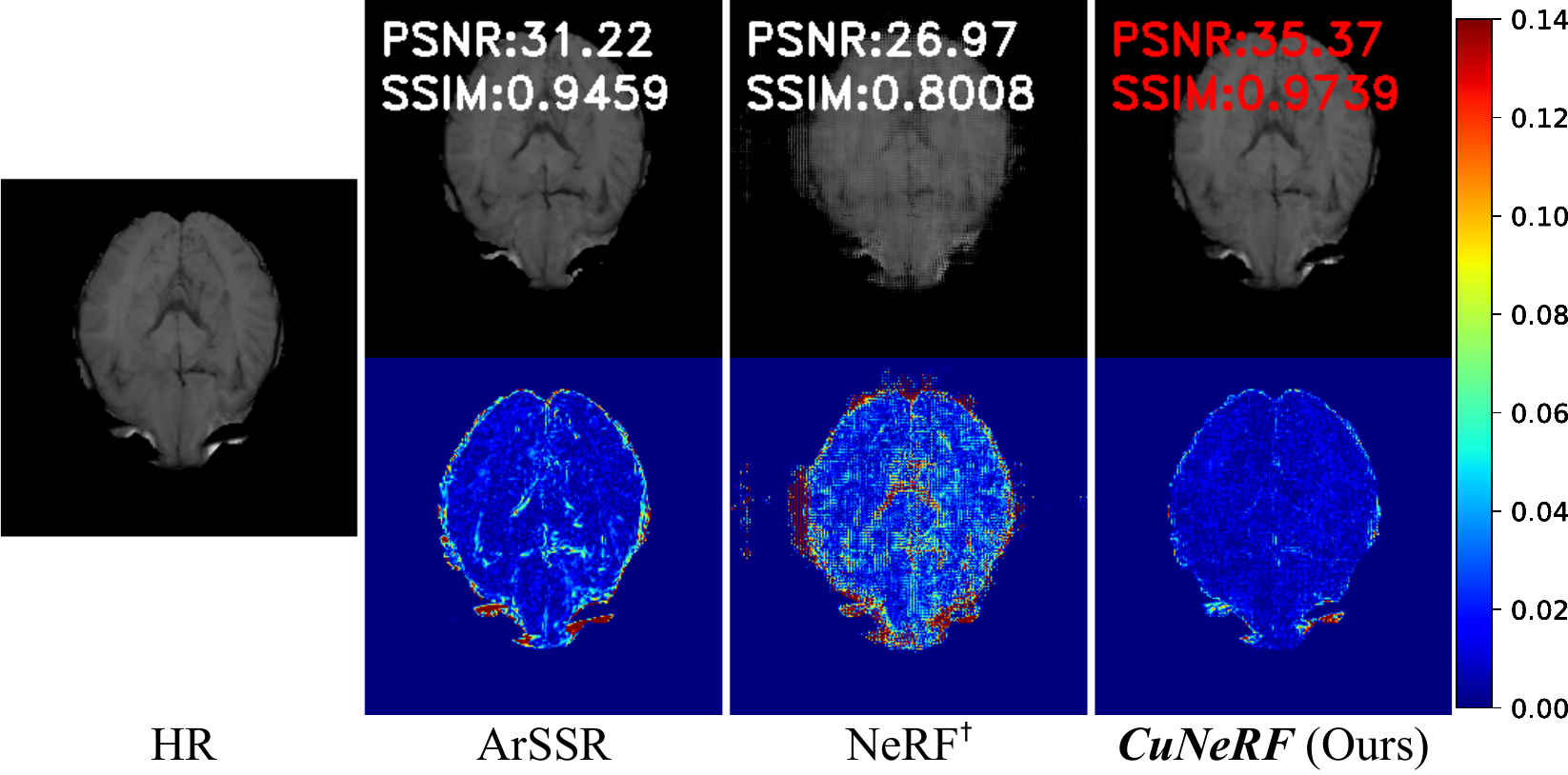}
  \caption{
    Visual examples of 3D MISR at $\times$2.5 factor between ArSSR \cite{ARSSR}, NeRF$^\dag$ \cite{NeRF} and our \textit{CuNeRF} on MSD \cite{MSD} dataset.
    Heatmaps at the bottom visualize the difference between the results and the HR image.
    Visually, NeRF$^\dag$ yields grid-like artifacts, and ArSSR produces non-existent details.
    By contrast, our \textit{CuNeRF} achieves better visual verisimilitude and fewer artifacts.
  }
  \label{fig:examples}
\end{figure}

\section{Method}
\begin{figure*}[!t]
  \centering
  \includegraphics[width=\textwidth]{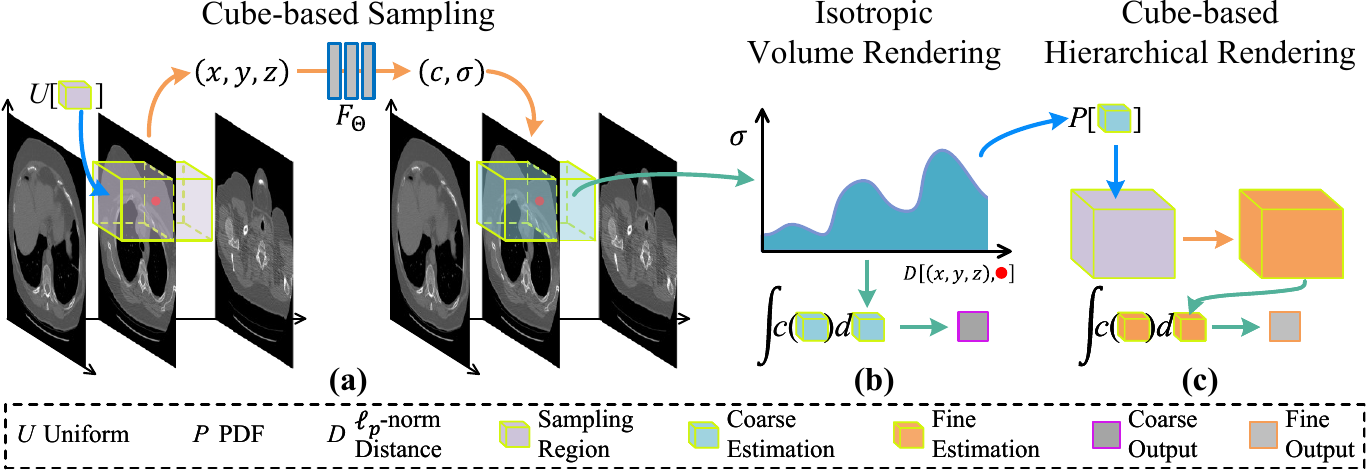}
  \caption{
    The overall framework of our \textit{CuNeRF}.
    To synthesize a pixel (\textcolor{red}{red circle}) with the spatial position $\mathbf{x}_t$, \bt{(a)} \textit{CuNeRF} first uniformly samples $N$ points as a point set $\{\hat{\mathbf{x}}_i\}^N_{i=1}$ within the cube space (\textcolor{Orchid}{purple cube}) centered by $\mathbf{x}_t$.
    Then, \textit{CuNeRF} obtains the coarse estimation (\textcolor{LightSkyBlue}{blue cube}) by feeding the sampling points into an MLP $F_{\Theta}$ to produce the set of corresponding pixel intensity $\{c_i\}_{i=1}^N$ and volume density $\{\sigma_i\}_{i=1}^N$.    
    \bt{(b)} Subsequently, assuming $\sigma$ of each sampling point is only related to the distance with the cube center $\mathbf{x}_t$, \textit{CuNeRF} computes the coarse output of the target pixel via volume integral.
    \bt{(c)} Finally, \textit{CuNeRF} resamples the points under the probability density function (PDF) of coarse estimation to acquire the fine estimation (\textcolor{BurntOrange}{orange cube}) of the cube.
    The fine output is generated by the same procedures as \bt{(b)}.
    Since these two rendering functions are differentiable, \textit{CuNeRF} can be optimized by minimizing the rendering loss in Eq \ref{eq:render_loss}.
    The fine output is the final rendering result of the target spatial position $\mathbf{x}_t$.
  }
  \label{fig:overall}
\end{figure*}

In this section, we first analyze the limitations of NeRF for rendering medical volumes and elaborate on our motivations.
Subsequently, based on our findings, we propose cube-based NeRF (CuNeRF), a novel yet efficient method to deal with ``zero-shot'' medical image arbitrary-scale super-resolution (MIASSR), which extends NeRF's application scenarios in the medical domain.
Specifically, we first normalize the medical volumes into the range of $[-1, 1]$ by volumetric normalization, and then train the model via proposed differentiable modules: cube-based sampling, isotropic volume rendering, and cube-based hierarchical rendering. 
During training, \textit{CuNeRF} is building a coordinate-intensity continuous function whose input is a 3D location $\mathbf{x}$$=$$(x,y,z)$ and the output is the corresponding pixel value $c$.
After optimization, \textit{CuNeRF} can predict pixels at any spatial position within the range.
As a result, \textit{CuNeRF} is capable to render medical slices at free viewpoints and arbitrary scales by feeding the corresponding plane equations. 
Figure \ref{fig:overall} depicts the overall framework of our \textit{CuNeRF}, and the subsequent techniques are described in the following subsections.

\subsection{Analysis \& Motivation} \label{sec:holes}
\begin{figure}[!t]
  \centering
  \includegraphics[width=3.2in]{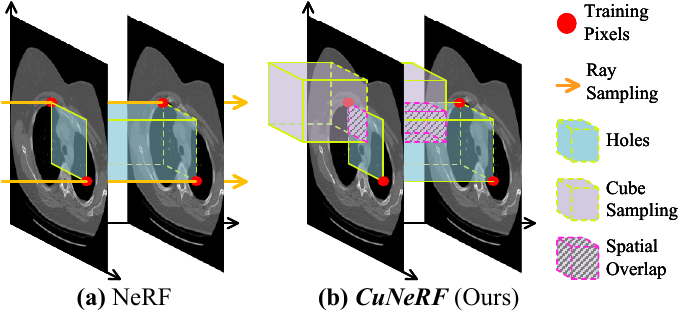}
  \caption{
    Visualization of the sampling strategies between NeRF \cite{NeRF} \bt{(a)} and \textit{CuNeRF} \bt{(b)} applied on medical volumes.
    Visually, NeRF \emph{only} samples the rays corresponding to each training pixel, which cannot cover the whole representation fields, leaving some ``holes'' (\emph{i.e.}, unmodeled space within between adjacent training pixels.
    To address this issue, \textit{CuNeRF} samples cubes centered by each training pixel, and therefore the ``holes'' are well-covered by the spatial overlaps.
  }
  \label{fig:motivation}
  \end{figure}

As shown in Figure \ref{fig:examples}, NeRF's sampling strategy may not be suitable for directly applying to medical volumes, which may produce grid-like artifacts in the results.
To explain this limitation, we provide an example of NeRF's modeling strategies applied to medical volumes in Figure \ref{fig:motivation} \bt{(a)}.
Visually, NeRF is trained to model the volumetric space along the ray emitted by each training pixel.
Since medical volumes only contain three orthogonal slices, which differs from multi-view photos collected by conventional cameras, and thus NeRF's modeling techniques cannot cover the entire representation fields, leaving some ``holes'' (\emph{i.e.,} unmodeled space) within the regions between adjacent training pixels.
Consequently, NeRF may produce sub-optimal results while rendering the contents within the holes.
As shown in Figure \ref{fig:examples}, NeRF$^\dag$\footnote{NeRF$^\dag$ is trained on three-orthogonal views.} produces grid-like artifacts in upsampling medical volumes, which demonstrates NeRF may struggle to render high-quality HR medical volumes from the corresponding LR ones.

To address the hole-forming issues caused by NeRF's ray sampling, we introduce cube-based sampling, which samples cubes (3D volumetric space) instead of rays (1D space) to fill the hole regions between adjacent training pixels by the spatial overlaps, as demonstrated in Figure \ref{fig:motivation} \bt{(b)}.
To adapt cube-based sampling, we further propose isotropic volume rendering and cube-based hierarchical rendering modules.
These modules will be introduced in the following subsections.

\subsection{Volumetric Normalization}
To build the continuous volumetric representations for the given medical volumes, we first normalize the whole volumetric space $H\!\times\!W\!\times\!L$ into an $\ell_{\infty}$ open ball as:
\begin{equation}
  \mathcal{B}(\hat{\mathbf{x}}_{\mathbf{o}}, 1)=\{\hat{\mathbf{x}}:\|\hat{\mathbf{x}}-\hat{\mathbf{x}}_{\mathbf{o}}\|_{\infty}<1\},
\end{equation}
where $\hat{\mathbf{x}}_{\mathbf{o}}$ is set to $(0,0,0)$ as the center $\mathbf{x}_{\mathbf{o}}=(\frac{H}{2}, \frac{W}{2}, \frac{L}{2})$ of the medical volume.
To adapt the positional encoding $\gamma(\cdot)$ introduced in Eq \ref{eq:PE}, each positional coordinate $\mathbf{x}_{t}=(x_{t},y_{t},z_{t})$ within the medical volume is transformed into the field coordinate $\hat{\mathbf{x}}_{t}=(\hat{x}_{t},\hat{y}_{t},\hat{z}_{t})$. 
The normalization function $\mathcal{N}(\cdot)$ is formulated as:
\begin{equation}
  \hat{\mathbf{x}}_{t} =\left(\frac{2(x_{t}-\frac{H}{2})}{H+2P}, \frac{2(y_{t}-\frac{W}{2})}{W+2P}, \frac{2(z_{t}-\frac{L}{2})}{L+2P}\right),
\end{equation}
where $P$ is a hyperparameter as the padding size.

\subsection{Cube-based Sampling}
Implicit neural representation methods aim to build the continuous representation of medical volumes.
However, NeRF suffers from hole-forming issues, which may leave some unmodeled spaces in their representation fields, and thus synthesizes grid-like artifacts in upsampled results.
To circumvent the holes forming in the representation fields, we propose a novel sampling strategy: cube-based sampling, which samples cubes (3D volumetric space) instead of rays (1D space).
Specifically, for the spatial position $\hat{\mathbf{x}}_{t}$, \textit{CuNeRF} samples a set of points within the cube space $\mathcal{B}(\hat{\mathbf{x}}_{t}, \frac{l}{2})$.
Each point $\hat{\mathbf{x}}_{i}$ is chosen under the uniform distribution $\mathcal{U}$ by:
\begin{equation}
  \hat{\mathbf{x}}_{i}\sim\mathcal{U}\left[\mathcal{B}(\hat{\mathbf{x}}_{t}, \frac{l}{2})\right], 
\end{equation}
where $l$ denotes the edge length of the cube.
We employ the group of these $N$ sampling points to approximate the cube space.
Due to the spatial overlaps between adjacent cubes, the representation fields can be well-covered by employing the proposed cube-based sampling in optimization.
As a result, the representation fields can be densely modeled with the same sampling time as NeRF \cite{NeRF}.

\subsection{Isotropic Volume Rendering}
As introduced in \ref{sec:NeRF}, the pixel color related to the ray $\mathbf{r}$ is computed by an integral in Eq \ref{eq:integral}.
Intuitively, the pixel color $\mathbf{C}(\hat{\mathbf{x}}_{t}, l)$ related to the cube space $\mathcal{B}(\hat{\mathbf{x}}_{t}, \frac{l}{2})$ can be computed by the following triple integral as:
\begin{equation}
  \label{eq:triple}
  \mathbf{C}(\hat{\mathbf{x}}_{t}, l)=\!\!\!\iiint\limits_{\mathcal{B}(\hat{\mathbf{x}}_{t}, \frac{l}{2})}\!\frac{\sigma(\hat{x},\hat{y},\hat{z})\mathbf{c}(\hat{x},\hat{y},\hat{z})d\hat{x}d\hat{y}d\hat{z}}{\exp(\int_{\hat{x}_{n}}^{\hat{x}}\!\int_{\hat{y}_{n}}^{\hat{y}}\!\int_{\hat{z}_{n}}^{\hat{z}}\!\sigma(x,y,z)dxdydz)},
\end{equation}
where $(\hat{x}_{n},\hat{y}_{n},\hat{z}_{n})=(\hat{x}_{t}-\frac{l}{2}, \hat{y}_{t}-\frac{l}{2}, \hat{z}_{t}-\frac{l}{2})$ denotes the initial location of the triple integral while $\mathbf{c}(\cdot)$ and $\sigma(\cdot)$ represent the color and volume density functions.
However, since NeRF samples $N$ points to approximate the volume rendering integral of the ray using numerical quadrature in Eq \ref{eq:volume}, it is required to sample $N^{3}$ points to model the cube with the same density, leading to massive computational costs.

Inspired by CRF \cite{crf} that assigns the nearby pixels with similar potentials, we assume the volume density $\sigma$ of each point $\hat{\mathbf{x}}$ within the cube $\mathcal{B}(\hat{\mathbf{x}}_{t}, \frac{l}{2})$ is only related to the $\ell_{p}$ distance $r=\|\hat{\mathbf{x}} - \hat{\mathbf{x}}_{t}\|_{p}$ between the centroid and itself.
Hence, the volumetric distribution of the cube is isotropic towards the value of $r$.
The above triple integral can be converted into the spherical coordinate system by:
\begin{equation}
  \label{eq:sim}
  \mathbf{C}(\hat{\mathbf{x}}_{t}, l) = 4\pi\int_{0}^{\hat{r}}\frac{r^{2}\sigma(\hat{\mathbf{x}}_{t}, r)c(\hat{\mathbf{x}}_{t}, r)dr}{\exp(4\pi\int_{0}^{r}s^{2}\sigma(\hat{\mathbf{x}}_{t}, s)ds)}, 
\end{equation}
where $\hat{r}=\|(\frac{l}{2},\frac{l}{2},\frac{l}{2})\|_{p}$ denotes the max distance of $r$ within the cube.
The derivation detail of Eq \ref{eq:sim} is shown in the supplementary materials.
Given $N$ sampling points by the proposed cube-based sampling, \textit{CuNeRF} first sorts these points by the distance $r$.
Subsequently, the integral of the cube is approximated via numerical quadrature rules:
\begin{equation}
  \hat{\mathbf{C}}(\hat{\mathbf{x}}_{t}, l)\!=\!4\pi\!\sum^{N}_{i=1}\frac{r_{i}^{2}(1-\exp(-\sigma_{i}(r_{i+1}\!-\!r_{i})))}{\exp(4\pi\sum_{j=1}^{i}r_{i}^{2}\sigma_{j}(r_{j+1}\!-\!r_{j}))}\mathbf{c}_{i},
  \label{eq:M_volume}
\end{equation}
where $\hat{\mathbf{C}}(\hat{\mathbf{x}}_{t}, l)$ denotes the predicted color of $\hat{\mathbf{x}}_{t}$.

\subsection{Cube-based Hierarchical Rendering}
To refine the results, \textit{CuNeRF} allocates sampling points proportionally to their expected volume distribution within the cube.
Similar to NeRF, \textit{CuNeRF} also simultaneously optimizes the coarse and fine MLPs. 
As obtaining the coarse output $\hat{\mathbf{C}}_{c}(\hat{\mathbf{x}}_{t}, l)$, \textit{CuNeRF} first samples $N_{f}$ numbers of $r$ using ITS.
Subsequently, for each $r$, we use the hierarchical sampling function $\zeta_{p}(\cdot)$ to select important points:
\begin{equation}
\hat{\mathbf{x}}_{f}=\zeta_{p}(r, \varphi, \theta),
\end{equation}
where $\varphi$ and $\theta$ are the randomly sampled spherical coordinates, and $\zeta_{p}(\cdot)$ converts the $\ell_{p}$ spherical coordinates $(r,\varphi,\theta)$ to the Cartesian coordinates $\hat{\mathbf{x}}$.
If $p\neq \infty$, we allow $\hat{\mathbf{x}}_{f}$ can beyond the cube space $\mathcal{B}(\hat{\mathbf{x}}_{t}, \frac{l}{2})$.
After obtaining fine outputs $\hat{\mathbf{C}}_{f}(\hat{\mathbf{x}}_{t}, l)$ at Eq \ref{eq:M_volume} using the sorted union of $N_{c}+N_{f}$ sampling points, \textit{CuNeRF} can be optimized in each batch $\mathcal{R}$ by the proposed adaptive rendering loss:
\begin{equation}
  \begin{split}
  \mathcal{L}_{A}\!=\!\!\!\!\!\sum_{\hat{\mathbf{x}}_{t}\in \mathcal{R}}\!\!\!\left[\lambda\|g.t. \!-\! \hat{\mathbf{C}}_{c}(\hat{\mathbf{x}}_{t}, l)\|^{2}_{2}\!+\!\|g.t. \!-\! \hat{\mathbf{C}}_{f}(\hat{\mathbf{x}}_{t}, l)\|^{2}_{2}\right]\!\!,
  \end{split}
  \label{eq:render_loss}
\end{equation}
where $\lambda=\|g.t. - \hat{\mathbf{C}}_{f}(\hat{\mathbf{x}}_{t}, l)\|^{\frac{1}{2}}$ is an adaptive regularization term to alleviate the overfitting brought by the ``coarse'' part.

\subsection{Medical Slice Synthesis}\label{sec:test}
After optimization, \textit{CuNeRF} can predict the pixels at any spatial coordinates within the representation fields.
Therefore, \textit{CuNeRF} can represent medical slices with free viewpoints and arbitrary scales by feeding the corresponding plane coordinates.
Detailed techniques are described in the following, and we show some examples in Section \ref{sec:rendering}.

\noindent\bt{Free-Viewpoint Rendering.} To render a medical slice with the given position $\hat{\mathbf{x}}$ and viewpoint $\mathbf{d}$, we first construct a base plane $\mathcal{P}_{o}$ at $\hat{\mathbf{x}}_{o}$.
Subsequently, we employ the translation matrix $\mathcal{M}_{T}$ to move slices from $\hat{\mathbf{x}}_{o}$ to $\hat{\mathbf{x}}$.
Finally, since the viewpoint $\mathbf{d}$ can be represented as rotating $\phi$ degrees around a certain axis $n_{\perp}$, we can obtain the rotation matrix $\mathcal{M}_{R}$ via Rodrigues' rotation formula \cite{rodrigues}.
Thus, the target plane $\mathcal{P}_t$ can be calculated as:
\begin{equation}
  \mathcal{P}_t=\mathcal{M}_{T}\mathcal{M}_{R}\mathcal{P}_{o}.
\end{equation}
The target medical slices can be obtained by feeding the points sampled within $\mathcal{P}_t$ into our \textit{CuNeRF}.

\noindent\bt{Arbitrary-Scale Rendering.} 
To render a medical slice with the given sampling scale $\delta$, we first follow the above process to obtain the target plane $\mathcal{P}_t$.
Then, we calculate the scale matrix $\mathcal{M}_{S}$ based on $\delta$, and sample the points under the translation $\mathcal{M}_{S}\mathcal{P}_t$.
By feeding the sampling points into our \textit{CuNeRF}, we can obtain the desired medical slices.

\begin{figure}[!t]
  \centering
  \includegraphics[width=3.2in]{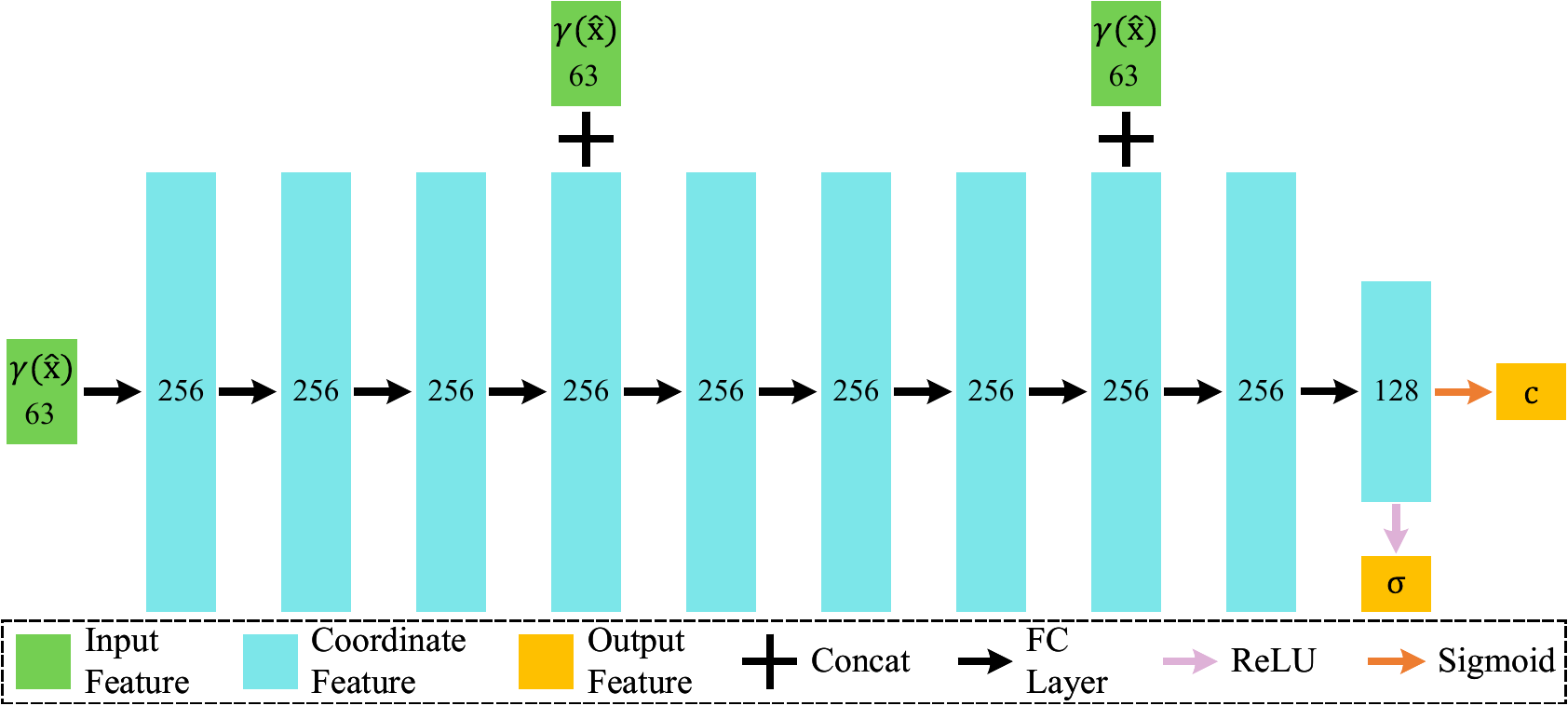}
  \caption{
    The architecture of MLP.
    For a given coordinate $\hat{\mathbf{x}}$, it is first encoded by $\gamma(\cdot)$ in Eq \ref{eq:PE} as input features.
    Then, we pass it through 9 fully-connected layers, each having $256$ channels with a ReLU.
    We concatenate the input features to the 4th and 8th hidden layers as the skip connections.
    Finally, we downscale the feature channels to predict a volume density $\sigma$ and pixel intensity $\mathbf{c}$.
  }
  \label{fig:MLP}
\end{figure}

\begin{table*}[!t]
  \setlength{\tabcolsep}{1.5mm}
  \caption{
    3D MISR comparisons on MSD \cite{MSD} dataset.
    \bt{Bold} and \ul{underline} texts indicate the best and second best performance.
  }
  \label{tab:3DMISR}
\centering
\footnotesize
\begin{tabular}{l|cc|cc|cc|cc|cc|cc|cc}
\hline
                         & \mc{$\times2$}          & \mc{$\times2.5$}        & \mc{$\times3$}           & \mc{$\times4$}          & \mc{$\times5$}          & \mc{$\times6$}          & \mc{$\times8$}           \\\cline{2-15}
                         & PSNR\ua    & SSIM\ua    & PSNR\ua    & SSIM\ua    & PSNR\ua    & SSIM\ua     & PSNR\ua    & SSIM\ua    & PSNR\ua    & SSIM\ua    & PSNR\ua    & SSIM\ua    & PSNR\ua    & SSIM\ua     \\\hline
\multicolumn{15}{c}{Conventional methods}\\\hline
Bicubic                  & 33.75      & 0.9469     & 30.84      & 0.9271     & 30.74      & 0.9161      & 28.67      & 0.8721     & 28.14      & \ul{0.8687}& 26.83      & 0.8521     & 26.54      & \ul{0.8376} \\\hline
\multicolumn{15}{c}{Supervised MIASSR methods}\\\hline
ArSSR \cite{ARSSR}       & \ul{36.98} & \ul{0.9690}& \ul{35.24} & \ul{0.9398}& \ul{34.69} & \ul{0.9199} & \ul{33.21} & \ul{0.8910}& \ul{30.50} & 0.8624     & \ul{29.96} & \ul{0.8543}& \ul{28.43} & 0.8353      \\\hline
\multicolumn{15}{c}{Zero-Shot MIASSR methods}\\\hline
NeRF$^{\dag}$ \cite{NeRF}& 29.33      & 0.8472     & 27.03      & 0.8392     & 25.98      & 0.8220      & 25.12      & 0.8088     & 24.50      & 0.7767     & 23.45      & 0.7549     & 22.63      & 0.7275      \\
\textbf{CuNeRF}          & \bt{39.62} & \bt{0.9786}& \bt{37.56} & \bt{0.9441}& \bt{36.24} & \bt{0.9267} & \bt{35.01} & \bt{0.9031}& \bt{34.73} & \bt{0.8952}& \bt{33.69} & \bt{0.8800}& \bt{31.19} & \bt{0.8675} \\\hline
\end{tabular}
\end{table*}

\section{Experiments}
In this section, we conduct extensive experiments and in-depth analysis to demonstrate the superiority of our \textit{CuNeRF} in representing high-quality medical images at arbitrary scales. 
For fair comparisons, the hyperparameters and model settings are consistent in all experiments.

\subsection{Experimental Details}\label{sec:details}
\noindent\bt{Datasets. }
We comprehensively compare our \textit{CuNeRF} and the existing advances in 2 different modalities: CT and MRI.
More specifically, we select T1-weighted MRI volumes from the Medical Segmentation Decathlon (MSD) \cite{MSD} while we also take CT volumes from the 2019 Kidney Tumor Segmentation Challenge (KiTS19) \cite{kits19} datasets, respectively.
All MRI volumes have the same dimension of $240$$\times$$240$$\times$$155$.
The image size of each CT slice is $512$$\times$$512$, while the number of CT slices is different.

In experiments, we resize all the CT volumes into $512^{3}$.
All the MRI and CT volumes are normalized into $[0, 1]$.
The degradation strategy is the nearest-neighbor interpolation.
\textit{For the compared supervised methods} \cite{ARSSR,saint,tvsrn}, we select 50 LR-HR MRI pairs to finetune the pre-trained model of \cite{ARSSR}, and also select 150 LR-HR CT pairs to train\cite{saint,tvsrn}.
The evaluation set consists of 80 medical volumes, including 40 CT and 40 MRI volumes.
\textit{Note}, following the ZSSR settings reported in \cite{zssr, survey}, we train NeRF$^{\dag}$ and our \textit{CuNeRF} on each LR test volume itself (see Figure \ref{fig:zssr}), while the HR volumes are only used for evaluations.

\noindent\bt{Multi-Layer Perceptron Architecture. }\label{sec:MLP}
Figure \ref{fig:MLP} depicts MLP's architecture, where the input is a 3D location $\hat{\mathbf{x}} = (\hat{x}, \hat{y}, \hat{z})$ encoded by $\gamma(\cdot)$ and the output is a 2D union of pixel intensity $\mathbf{c}$ and volume density $\sigma$.
The parameter size of the proposed MLP is about 0.58M

\noindent\bt{Implementation Details. }
\textit{CuNeRF} is implemented on top of \cite{nerf-pytorch}, a Pytorch \cite{Pytorch} re-implementation of NeRF.
Our experiments run on a single NVIDIA RTX 3090 GPU with 24G memory.
We employ the $\ell_{2}$ distance for isotropic volume rendering and hierarchical cubic rendering while the edge length $l$ of the cube is set to $1$.
The hierarchical sampling function $\zeta_{2}(\cdot)$ converts $(r, \varphi, \theta)$ to $\hat{\mathbf{x}}\!=\!(\hat{x}, \hat{y}, \hat{z})$ by:
\begin{equation}
    \begin{split}
      \hat{\mathbf{x}}=(r\sin{\varphi}\cos{\theta}, r\sin{\varphi}\sin{\theta}, r\cos{\varphi}),
    \end{split}
\end{equation}
where $\varphi\sim\mathcal{U}[0, \pi]$ and $\theta\sim\mathcal{U}[0, 2\pi]$, respectively.
Let $p=2$ as default, we consider a spherical parameterized as $(\hat{x}, \hat{y}, \hat{z})=\zeta_{2}(r, \varphi, \theta)$, where $\varphi \in [0, \pi], \theta \in [0, 2\pi],r>0$.
This change of variables from the Cartesian system gives us a differential term:
\begin{align}
  d\hat{x}d\hat{y}d\hat{z}\!&=\!|det(D\zeta_{2})|drd\varphi d\theta\\
    &=\!r^{2}\sin\varphi drd\varphi d\theta.
\end{align}
Therefore, the volume rendering function in Eq \ref{eq:sim} can be simplified from Eq \ref{eq:triple} as follow:
\begin{align}
  \mathbf{C}(\hat{\mathbf{x}}_{t}, l)\!&=\!\!\!\int_{0}^{2\pi}\!\!\!\!\int_{0}^{\pi}\!\!\!\int_{0}^{\frac{\sqrt{3}}{2}l}\!\!\!\frac{\sigma(\hat{\mathbf{x}}_{t}, r)\mathbf{c}(\hat{\mathbf{x}}_{t}, r)r^{2}\sin\varphi d\theta d\varphi dr}{\exp(\!\!\!\iiint\limits_{\mathcal{B}(\hat{\mathbf{x}}_{t}, r)}\!\!\!\sigma(\hat{\mathbf{x}}_{t}, s)s^{2}\!\sin\varphi^{'}\!d\theta^{'}\!d\varphi^{'}\!ds)}\\
  &=4\pi\int_{0}^{\frac{\sqrt{3}}{2}l}\frac{\sigma(\hat{\mathbf{x}}_{t}, r)c(\hat{\mathbf{x}}_{t}, r)r^{2}dr}{\exp(4\pi\int_{0}^{r}\sigma(\hat{\mathbf{x}}_{t}, s)s^{2}ds)}. 
\end{align}

\textit{For training}, we employ Adam \cite{adam} as the optimizer with a weight decay of $10^{-6}$ and a batch size of $2048$.
The maximum iteration is set to $250000$, and the learning rate is annealed logarithmically from $2$$\times$$10^{-3}$ to $2$$\times$$10^{-5}$.
Similar to NeRF, \textit{CuNeRF} first samples $64$ points for the coarse MLP $F^{c}_{\Theta}$ and feeds $192$ points (the sorted union of $64$ coarse and $128$ fine points) into the fine MLP $F^{f}_{\Theta}$.
The training time for each $512$$\times$$512$$\times$$512$ volume is about $0.8$$\sim$$3$ \textit{hours}.

\textit{For testing}, the number of sampling points is set to $16$ ($8$ for coarse MLP and $8$ for fine MLP), which can reduce considerable computational costs.
The results are obtained by feeding all the coordinates of the given plane equations into our model (seeing details in Section \ref{sec:test}).
The inference time for rendering a $256$$\times$$256$$\times$$256$ volume is about 30 \textit{secs}.
\emph{Note} that we do not use any pre- and post-processing techniques to improve our results in the experiments.

\noindent\bt{Evaluation Metrics. }
We use two quantitative metrics: Peak Signal-to-Noise Ratio (PSNR) and Structured Similarity Index (SSIM) \cite{ssim} to measure the image quality of different methods.
Note we report the average SSIM on axial, coronal, and sagittal planes for volumetric MISR.

\begin{table}[!t]
  \setlength{\tabcolsep}{1.2mm}
  \caption{
    Quantitative comparisons of start-of-the-art methods on KiTS19 \cite{kits19} dataset for volumetric MISR.
    \bt{Bold} and \ul{underline} texts indicate the best and second best performance.
  }
  \label{tab:ZMISR}
\centering
\footnotesize
\begin{tabular}{l|cc|cc|cc}
\hline
                         & \mc{$\times2$}           & \mc{$\times4$}    & \mc{$\times8$}  \\\cline{2-7}
                         & PSNR\ua    & SSIM\ua     & PSNR\ua    & SSIM\ua     & PSNR\ua    & SSIM\ua \\\hline     
\multicolumn{7}{c}{Conventional methods}\\\hline
Bicubic                  & 37.75      & 0.9498      & 33.76      & 0.9149      & 29.03      & 0.8572\\\hline
\multicolumn{7}{c}{Supervised MISR methods}\\\hline
TVSRN \cite{tvsrn}       & \ul{39.32} & \bt{0.9790} & \ul{36.62} & \ul{0.9532} & \ul{32.10} & 0.9163 \\\hline
\multicolumn{7}{c}{Supervised MIASSR methods}\\\hline
SAINT \cite{saint}       & \bt{39.47} & \ul{0.9782} & 36.61      & \bt{0.9574} & 31.78      & \ul{0.9188}\\\hline
\multicolumn{7}{c}{Zero-Shot MIASSR methods}\\\hline
NeRF$^{\dag}$ \cite{NeRF}& 36.50      & 0.9383      & 34.14      & 0.9181      & 30.56      & 0.8748  \\
\textbf{CuNeRF}          & 38.33      & 0.9663      & \bt{36.64} & 0.9480      & \bt{32.44} & \bt{0.9216}   \\\hline
\end{tabular}
\end{table}

\begin{figure*}[!t]
  \centering
  \includegraphics[width=\textwidth]{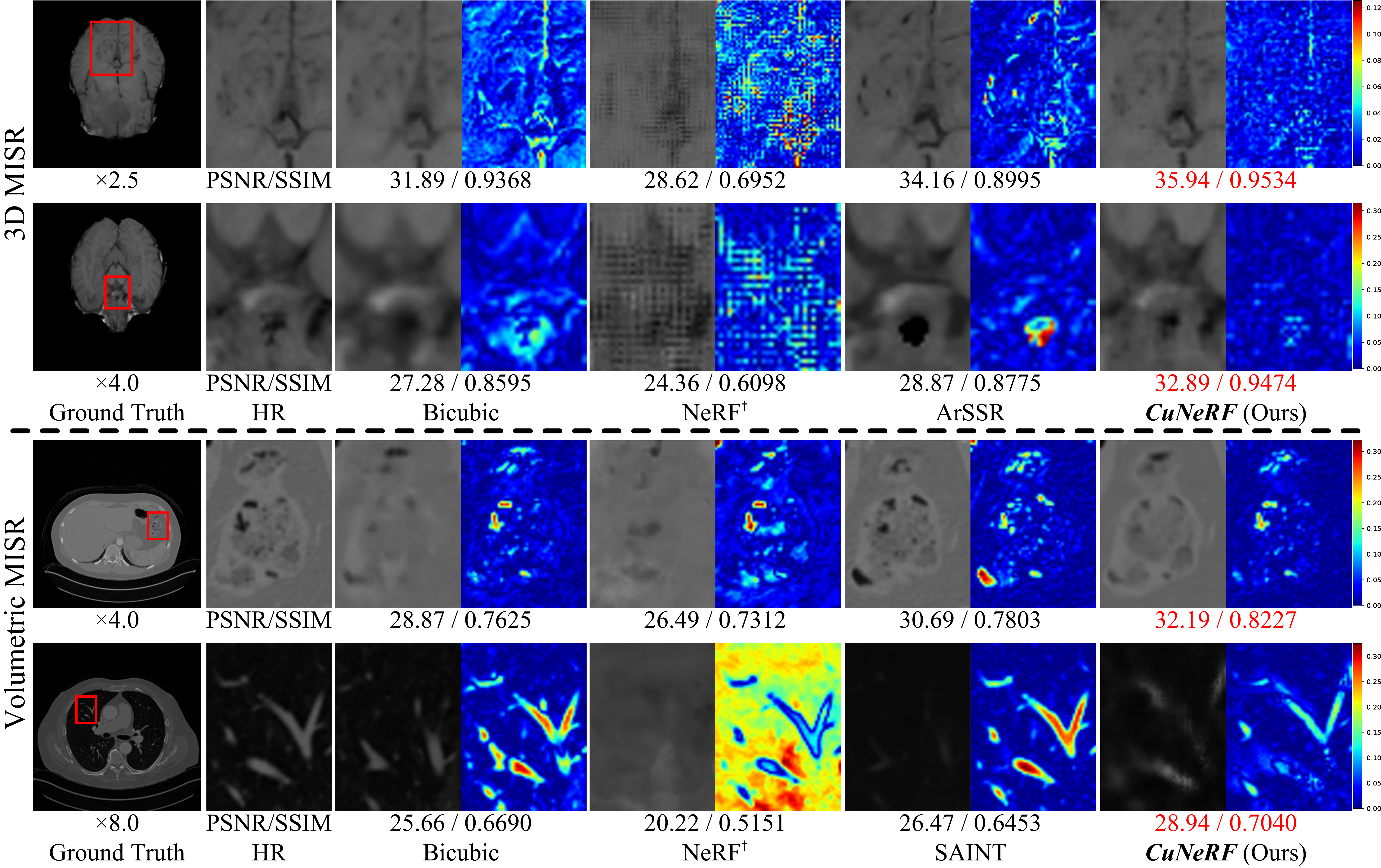}
  \caption{
    Visual comparisons between our \textit{CuNeRF} and 4 state-of-the-art methods: Bicubic, NeRF$^\dag$ \cite{NeRF}, ArSSR \cite{ARSSR} and SAINT \cite{saint} for 3D MISR and volumetric MISR.
    The heatmaps on the right of the results visualize the difference related to HR patches.
  }
  \label{fig:visual}
\end{figure*}

\subsection{Experimental Results}
We compare the proposed \textit{CuNeRF} with 5 state-of-the-art methods, including 2 supervised MIASSR methods: ArSSR \cite{ARSSR} and SAINT \cite{saint}, 1 supervised MISR method: TVSRN \cite{tvsrn}, 1 conventional method: bicubic interpolation, and NeRF$^{\dag}$ \cite{NeRF}.
Given a upsampling scale $\delta$, we evaluate these methods under the following two settings: \textit{(i)} \bt{3D MISR.} Upsampling the donwsampled volume from $\frac{H}{\delta}$$\times$$\frac{W}{\delta}$$\times$$\frac{L}{\delta}$ to $H$$\times$$W$$\times$$L$; \textit{(ii)} \bt{volumetric MISR.} Upsampling the donwsampled volume from $H$$\times$$W$$\times$$\frac{L}{\delta}$ to $H$$\times$$W$$\times$$L$.
\textit{Note} that NeRF$^{\dag}$ and \textit{CuNeRF} are trained with the same settings and similar parameter size ($\pm$0.02M).

\noindent\bt{Quantitative Comparison. }
We report 3D MISR and volumetric MISR based on the increasing upsampled scales in Table \ref{tab:3DMISR} and Table \ref{tab:ZMISR}, respectively.
As demonstrated, for the 3D MISR challenge on MRI volumes, \textit{CuNeRF} surpasses all the competitors with a consistent preferable performance at various upsampling scales.
For volumetric MISR challenge on CT volumes, \textit{CuNeRF} achieves comparable performance to SAINT \cite{saint} and TVSRN \cite{tvsrn}.
Compared to fully-supervised MIASSR methods: ArSSR \cite{ARSSR} and SAINT \cite{saint}, our \textit{CuNeRF} is more robust at presenting large-scale medical slices and capable to deal with different modalities (CT and MRI), suggesting \textit{CuNeRF} owns broader application scenarios.
It is also worth noting that NeRF$^{\dag}$ achieves comparable performance for volumetric MISR but fails in 3D MISR.
Since volumetric MISR only aims to acquire the pixels along the $z$-axis, the experimental results of NeRF$^{\dag}$ confirm our motivations.

\noindent\bt{Visual Comparison. }
We visualize the rendering results of \textit{CuNeRF} and other competitors on MRI (rows 1 and 2) and CT (rows 3 and 4) modalities in Figure \ref{fig:visual}.
It can be observed that \textit{CuNeRF} well represents the medical slices at various scales.
Compared to the exhibited methods, \textit{CuNeRF} is most similar to the ground truths, achieving better visual verisimilitude and reducing aliasing artifacts, especially in representing large-scale medical slices.
Since NeRF$^{\dag}$ exhibits grid-like artifacts in rendering high-quality medical slices at larger-valued scales, the visualization results prove the effectiveness of \textit{CuNeRF}, which extends NeRF's capability to continuously represent medical images.

\begin{figure*}[!t]
  \centering
  \includegraphics[width=\textwidth]{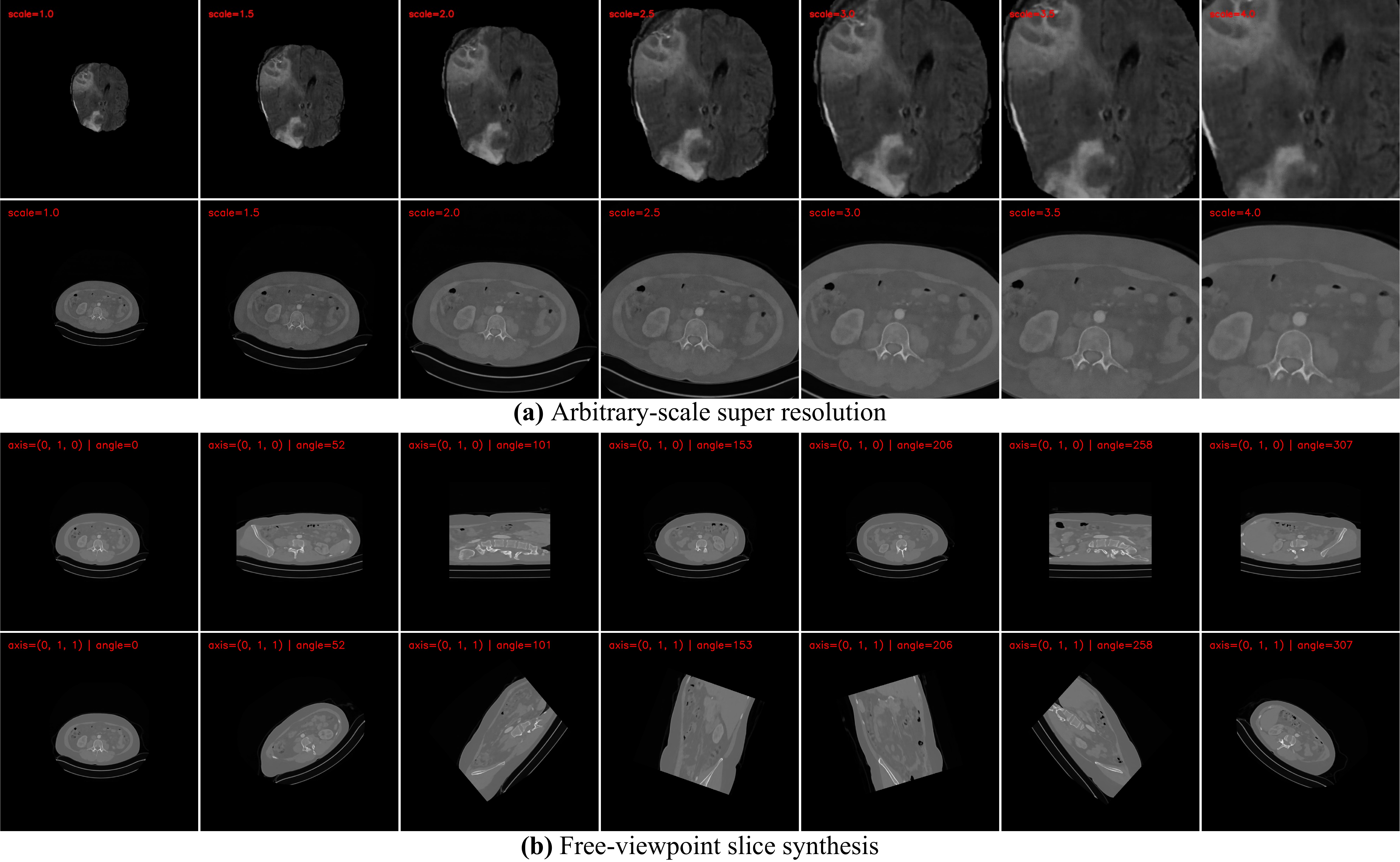}
  \caption{
    Visualization results at arbitrary scales \bt{(a)} and free viewpoints \bt{(b)} within a 1024$\times$1024 range.
  }
  \label{fig:rendering}
\end{figure*}

\noindent\bt{Free-Viewpoint \& Arbitrary-Scale Rendering. }\label{sec:rendering}
As shown in Figure \ref{fig:rendering}, \textit{CuNeRF} can synthesize medical images at continuous-valued scales \bt{(a)}.
Moreover, \textit{CuNeRF} is capable to yield medical slices with a viewpoint rotating 360 degrees around an arbitrary coordinate axis $n_{\perp}$.
Compared to existing methods, \textit{CuNeRF} is capable to provide richer visual information for clinical diagnosis.

\subsection{Ablation Study}
In this subsection, we conduct comprehensive experiments to prove the correctness of \textit{CuNeRF}'s design.
We first carry out ablation studies to investigate the effectiveness of the proposed modules.
Subsequently, we evaluate the \textit{CuNeRF}'s performance under different settings.

\noindent\bt{\textit{CuNeRF}'s ablation variants. }
We evaluate against several ablations of the proposed \textit{CuNeRF} with each module: \textit{CuS}, \textit{IVR} and $\mathcal{L}_{A}$ represent cube-based sampling, isotropic volume rendering, and adaptive rendering loss, respectively.
The baseline model here is NeRF$^\dag$.
As reported in Table \ref{tab:abla}, the baseline model struggles to deal with 3D MISR issues (row 1), while adopting \textit{CuS} instead of ray sampling can significantly improve the performance (row 2).
Compared to NeRF's volume rendering function, employing \textit{IVR} (row 3) can further improve the slice synthesis quality, suggesting \textit{IVR} can better estimate the volumetric distribution, reducing aliasing artifacts raised by undersampling.
Since the coarse term of NeRF's rendering loss may affect the optimization, $\mathcal{L}_{A}$ (row 4) is able to alleviate this distraction.

\begin{table}[!t]
  \setlength{\tabcolsep}{1.1mm}
  \caption{
    Comparisons of ablation variants on MSD \cite{MSD} dataset for 3D MISR.
    \bt{Bold} text indicates the best performance.
  }
\label{tab:abla}
\centering
\footnotesize
\begin{tabular}{ccc|cc|cc|cc}
\hline
\mr{\textit{CuS}} & \mr{\textit{IVR}} & \mr{$\mathcal{L}_{A}$} & \mc{$\times2$}          & \mc{$\times4$}          & \mc{$\times8$}               \\\cline{4-9}
                  &                   &                        & PSNR\ua    & SSIM\ua     & PSNR\ua    & SSIM\ua     & PSNR\ua    & SSIM\ua       \\\hline
                  &                   &                        & 29.33      & 0.8472      & 25.12      & 0.8088      & 22.63      & 0.7275        \\
\ck               &                   &                        & 35.82      & 0.9244      & 33.29      & 0.8703      & 27.77      & 0.8398        \\
\ck               &\ck                &                        & 38.15      & 0.9524      & 34.27      & 0.8887      & 29.34      & 0.8455        \\
\ck               &\ck                &\ck                     & \bt{39.62} & \bt{0.9786} & \bt{35.01} & \bt{0.9031} & \bt{31.19} & \bt{0.8675}   \\\hline
\end{tabular}
\end{table}

\noindent\bt{\textit{CuNeRF} under different settings. }
We evaluate the performance of \textit{CuNeRF} under different settings: ``$p$$=$$\infty$'' employs $\ell_{\infty}$ distance of $r$, ``$l$$=$$0.5$'' and ``$l$$=$$2$'' represent to set the edge length to $0.5$ and $2$ pixel distance, respectively.
The default is introduced in Section \ref{sec:details}, where $p$$=$$2$ and $l$$=$$1$.
As reported in Table \ref{tab:diff}, the default setting of \textit{CuNeRF} achieves consistent outperformance at various scales.
In contrast, employing the $\ell_{\infty}$ is significantly inferior to default, which means $\ell_{2}$ distance is more suitable to model the continuous representation for medical volumes.
Meanwhile, different cube edge $l$ acquire comparable performance to default, suggesting our \textit{CuNeRF} is a parameter-insensitive method with good robustness under different experimental settings.

\begin{table}[!t]
  \setlength{\tabcolsep}{1.0mm}
  \caption{
    Quantitative comparisons of \textit{CuNeRF} under different settings on MSD \cite{MSD} dataset for 3D MISR. 
    \bt{Bold} and \ul{underline} texts indicate the best and second best performance.
  }
  \label{tab:diff}
  \centering
  \footnotesize
  \begin{tabular}{l|cc|cc|cc}
  \hline
                              & \mc{$\times2$}           & \mc{$\times4$}           & \mc{$\times8$}           \\\cline{2-7}
                              & PSNR\ua    & SSIM\ua     & PSNR\ua    & SSIM\ua     & PSNR\ua    & SSIM\ua     \\\hline
  \textit{CuNeRF} default.    & \ul{39.62} & \bt{0.9786} & \bt{35.01} & \bt{0.9031} & \ul{31.19} & \ul{0.8675} \\
  \textit{CuNeRF} $p=\infty$  & 35.78      & 0.9348      & 32.85      & 0.8853      & 29.25      & 0.8433      \\
  \textit{CuNeRF} $l=0.5$     & 38.17      & 0.9621      & \ul{35.00} & 0.9018      & \bt{31.53} & \bt{0.8704} \\
  \textit{CuNeRF} $l=2$       & \bt{39.65} & \ul{0.9723} & 34.57      & \ul{0.9011} & 30.85      & 0.8608      \\
  \hline

  \end{tabular}
\end{table}

\section{Conclusion}
In this paper, we present Cube-based NeRF (CuNeRF), a zero-shot framework for medical image arbitrary-scale super-resolution (MIASSR).
Instead of learning the mapping between LR-HR pairs, \textit{CuNeRF} learns the continuous volumetric representation from LR volumes, thus a well-trained model can yield medical images at arbitrary viewpoints and scales in a continuous domain.
Extensive experiments demonstrate that \textit{CuNeRF} outperforms state-of-the-art methods, yielding better visual effects and reducing artifacts at various upsampling factors.

\section*{Acknowledgement}
This project is supported by the Natural Science Foundation of China (No. 62072482).

{\small
\bibliographystyle{ieee_fullname}
\bibliography{egbib}
}

\end{document}